\documentclass[physics,article,submit,pdflatex,moreauthors]{Definitions/mdpi} 

\usepackage{amsmath}
\usepackage{siunitx}
\usepackage{chemformula}
\usepackage{bbm}

\newcommand{\GeV}{\si{\giga\electronvolt}}
\newcommand{\euro}{\text{\ EUR}}
\newcommand{\europercm}{\si{\euro \per \centi\m}}

\firstpage{1} 
\pubvolume{1}
\issuenum{1}
\articlenumber{0}
\pubyear{2024}
\copyrightyear{2024}
\datereceived{ } 
\daterevised{ } % Comment out if no revised date
\dateaccepted{ } 
\datepublished{ } 
\hreflink{https://doi.org/} % If needed use \linebreak
\usepackage{subcaption} % For subfigures
\Title{End-to-End Detector Optimization with Diffusion models: A Case Study in Sampling Calorimeters}

\TitleCitation{Title}

\Author{
    Kylian Schmidt $^{1*}$\orcidK{},
    Krishna Nikhil Kota$^1$\orcidZ{},
    Jan Kieseler$^{1,10}$\orcidI{},
    Andrea De Vita$^{2,3}$\orcidD{},
    Markus Klute $^1$\orcidY{},
    Abhishek$^{4}$\orcidB{},
    Max Aehle $^{5,10}$\orcidE{},
    Muhammad Awais $^{2,6}$\orcidF{},
    Alessandro Breccia$^2$\orcidP{},
    Riccardo Carroccio $^2$\orcidC{},
    Long Chen $^{5,10}$\orcidL{},
    Tommaso Dorigo $^{6,7,3,10}$\orcidT{},
    Nicolas R.\ Gauger $^{5,10}$\orcidG{},
    Enrico Lupi $^{2,3}$\orcidX{},
    Federico Nardi $^{2,8,10}$ \orcidN{}
    Xuan Tung Nguyen $^{3,5}$\orcidH{},
    Fredrik Sandin $^{6,10}$\orcidS{},
    Joseph Willmore$^{2}$\orcidO{},
    Pietro Vischia $^{9,10}$ \orcidA{}
}

\address{ %
\noindent
$^{1}$ \quad Karlsruhe Institute of Technology, 76131 Karlsruhe, Germany \\
$^{2}$ \quad Universit\`a di Padova, dipartimento di Fisica e Astronomia, Via F. Marzolo 8, 35131 Padova, Italy \\
$^{3}$ \quad INFN, sezione di Padova - Via F. Marzolo 8, 35131 Padova, Italy \\
$^{4}$ \quad National Institute of Science Education and Research, Jatni, 752050, India \\ 
$^{5}$ \quad University of Kaiserslautern-Landau (RPTU), Gottlieb-Daimler-Straße, 67663 Kaiserslautern, Germany \\
$^{6}$ \quad Lule{\aa} University of Technology, 971 87 Lule\aa, Sweden \\
$^{7}$ \quad Universal Scientific Education and Research Network, Italy \\
$^{8}$ \quad Laboratoire de Physique Clermont Auvergne, 63170 Aubière, France \\
$^{9}$ \quad Universidad de Oviedo and ICTEA, Spain \\
$^{10}$ \quad MODE Collaboration, \url{https://mode-collaboration.github.io}\\
$^*$ \quad Correspondence: \texttt{kylian.schmidt@kit.edu} \\
}
\abstract{
Recent advances in machine learning have opened new avenues for optimizing detector designs in high-energy physics, where the complex interplay of geometry, materials, and physics processes has traditionally posed a significant challenge. In this work, we introduce the \emph{end-to-end} AI Detector Optimization framework (AIDO) that leverages a diffusion model as a surrogate for the full simulation and reconstruction chain, enabling gradient-based design exploration in both continuous and discrete parameter spaces. Although this framework is applicable to a broad range of detectors, we illustrate its power using the specific example of a sampling calorimeter, focusing on charged pions and photons as representative incident particles. Our results demonstrate that the diffusion model effectively captures critical performance metrics for calorimeter design, guiding the automatic search for layer arrangement and material composition that aligns with known calorimeter principles. The success of this proof-of-concept study provides a foundation for future applications of end-to-end optimization to more complex detector systems, offering a promising path toward systematically exploring the vast design space in next-generation experiments.
}

\AuthorNames{Kylian Schmidt, Krishna Nikhil Kota, Jan Kieseler, Andrea De Vita, Abhishek, Max Aehle, Muhammad Awais, Alessandro Breccia, Riccardo Carroccio, Long Chen, Tommaso Dorigo, Nicolas R.\ Gauger, Enrico Lupi, Federico Nardi, Xuan Tung Nguyen, Fredrik Sandin, Joseph Willmore, Pietro Vischia}

\AuthorCitation{Schmidt, K.; Kota, K. N.; Kieseler, J.; De Vita, A.; Abhishek; Aehle, M.; Awais, M.; Breccia, A; Carroccio, R.; Chen, L.; Dorigo, T.; Gauger, N. R.; Lupi, E.; Nardi, F.; Nguyen, X. T.; Sandin, F.; Willmore, J.; Vischia, P.}

\begin{document}
%%%%%%%%%%%%%%%%%%%%%%%%%%%%%%%%%%%%%%%%%%

\section {Introduction \label{s:intro}}

Designing a high-performance detector for particle physics is inherently a high-dimensional optimization challenge, requiring the reconciliation of multiple objectives such as energy or momentum resolution, timing accuracy, and cost. The design space may include both continuous parameters ({\it e.g.}, layer thickness) and discrete parameters ({\it e.g.}, material choices) as well as the reconstruction algorithms that translate raw detector signals into physics observables. 
This complexity is further compounded by the stochastic nature of particle interactions, which complicates the application of gradient-based methods~\cite{Dorigo:2023}.
While sufficiently generic and differentiable reconstruction algorithms exist ({\it e.g.}, Refs.~\cite{Qasim:2019otl,Kieseler:2020wcq,Kortus:2025,Reuter:2024kja}), the high-fidelity simulations describing particle interactions with matter are typically not differentiable, thereby preventing a straightforward use of gradient descent.
Ongoing efforts to overcome this limitation include making established simulation packages directly differentiable~\cite{Aehle:2022a,Aehle:2022b,Aehle:2024c}, developing custom differentiable simulation pipelines from the ground up ({\it e.g.}, TomOpt~\cite{Strong:2023oew}), or employing local surrogate models valid within a limited trust region and retraining them once that region is exited~\cite{Shirobokov:2020tjt}. Each of these approaches faces its specific challenges, especially when dealing with detectors containing hundreds of thousands of readout channels.

To address these difficulties, an \emph{end-to-end} approach can be employed, whereby only one global performance quantity is predicted by a surrogate neural model for each sample in the dataset (event).
This stands in contrast with approaches which use generative models to recreate the statistic patterns of low-level information in the detector, such as~\cite{favaro_calodream_2024,mikuni_caloscore_2024,amram_denoising_2023,hashemi_deep_2024,krause_calochallenge_2024}.
The end-to-end strategy abstracts the problem away from the microscopic details of the simulation and reconstruction, permitting the integration of the model into a comprehensive software pipeline. 
Such an end-to-end surrogate has the advantage of capturing the essential mapping from design parameters to physics-relevant performance metrics without having to explicitly model each individual hit in a high-granularity detector.
It also enables more direct optimization loops, as gradients obtained from the end-to-end model can steer design choices directly. Notably, discrete parameters such as the presence or absence of specific sub-detectors or the choice of different materials can be accommodated by neural networks through careful encoding, or with techniques that effectively smooth the discrete design space~\cite{Strong:2023oew}.

Among the various detector subsystems in modern particle physics experiments, calorimeters stand out both for their critical role in measuring particle energies and for their inherently large design space. Calorimeters have been widely used since the 1950s to determine the energy of incident particles by sandwiching high-density passive layers and active materials that produce measurable signals.
In collider-based experiments, the performance demands on calorimeters have rapidly evolved to address new challenges: for instance, highly segmented calorimeters in both transverse and longitudinal directions can identify the hadronic decays of boosted heavy particles inside wide jets~\cite{PhysRevD.101.056019, dreyerlund,ATL-PHYS-PUB-2023-021,kasieczka2018}, and the availability of a large magnetic field combined with high segmentation permits efficient particle flow algorithms~\cite{CMS:2017yfk,Aaboud:2257597} to significantly improve event reconstruction.
Meanwhile, new fabrication and sensor technologies, such as 3D-printed scintillators or the use of highly granular silicon as the active medium~\cite{HGCAL-TDR}, further expand the design space and underscore the importance of systematic optimization.

In this work, we illustrate how the end-to-end framework AIDO can be deployed to optimize both continuous and discrete parameters of a sampling calorimeter. We focus on a simplified setup involving only photons and charged pions as incident particles, and we seek to identify the arrangement of passive and active layers that provides the best energy reconstruction performance. Far from claiming to discover new, cutting-edge calorimeter designs, our primary goal is to demonstrate that the pipeline can identify designs that align with well-known calorimeter principles. This validation paves the way for future studies involving more complex objectives and design choices, where human intuition alone may fall short of identifying optimal solutions.

This paper is organized as follows. In Sec.~\ref{sec:mapping_detector_performance}, we motivate the comparison of different detector designs based on a single metric computed from simulations. Section~\ref{sec:digital_twin_approach} introduces the concept of a digital twin and explains how it boosts gradient computation, making end-to-end optimization feasible. Section~\ref{sec:workflow} details the workflow structure for the iterative simulation-reconstruction-optimization loop, while Sec.~\ref{sec:discrete_parameters} describes how discrete detector parameters can be encoded and learned. In Sec.~\ref{sec:application_to_sampling_calo}, we apply this end-to-end optimization approach to a sampling calorimeter. Finally, Sec.~\ref{sec:conclusion} summarizes our findings and outlines potential applications and improvements for future work.

\section{Mapping of detector parameters to performance} \label{sec:mapping_detector_performance}

Geant4~\cite{GEANT4:2002zbu,Allison:2006ve,Allison:2016lfl} is a fast simulation framework that accurately reproduces the interaction of particles with matter, and is widely used in High Energy Physics (HEP) to understand the behavior of detectors.
The framework provides a common core of functionalities, on top of which developers can build their simulation software for their specific use-case.
The flexibility provided by Geant4 allows physicists and engineers to quickly test and refine the design of new detectors by sharing some tunable parameters with other programs (macro files, python bindings, or Command Line arguments).
While these interfaces are convenient for human users, they also open up another potential application if connected with a hyper-network, as discussed in Sec.~\ref{sec:digital_twin_approach}.

Typically, the outputs of a Geant4 simulation are divided into events, each representing the full initial particles to final readout chain, including for example the energy deposited in a single cell, the number of particles that passed through a given layer, or the direction of flight of a particle.
This low-level information is fed into custom algorithms which compute further physical quantities of interest, such as the momentum, the total recorded energy or the flight time.
One considerable advantage of simulations in general is that, in contrast to real HEP experiments, the initial conditions of an event are precisely known and reproducible.
Consequently, reconstruction algorithms that produce high-level variables, such as the energy resolution of a calorimeter or the accuracy of a tracking component, can be evaluated in an unbiased way.
It is this observation that motivates the automatic optimization of detectors, which can be done by tuning the parameters and computing the resulting performance with a reconstruction algorithm.

Given a detector simulation with a set of adjustable parameters $\theta$, we denote the configuration space of all possible detectors as $\Theta = \{ \theta^0 \times \theta^1 ... \times \theta^P\}$, where each parameter $\theta^p$ describes a certain physical property of the detector.
This includes continuous parameters, such as the position or thickness of a component, and categorical parameters, such as a material type.
For continuous parameters, one must ensure that the physical meaning is properly encoded by setting boundaries, {\it e.g.} forbidding negative thicknesses.
Further information about the translation of discrete parameters into a suitable format for Machine Learning (ML) is detailed in Sec.~\ref{sec:discrete_parameters}.

In a first step, we consider the simulation to be equivalent with a mapping $f:\theta, \theta_\text{ext} \mapsto F_\text{sim}$ that takes as inputs $\theta$ and a second set of external variables $\theta_\text{ext}$ which are potentially pseudo-random, making $f$ a stochastic process.
We categorize the outputs into three terms, $F_\text{sim} = \{x, T, C \}$, where $x$ is the set of all low-level, hit-based features (such as the energy deposited in each cell, the position of an individual hit, etc.).
The targets $T$, or true information, describe the known initial conditions of the simulation and are used for the evaluation of the reconstruction algorithm (such as the true initial energy).
Lastly, $C$ denotes all additional context information for each event (such as the initial particle type), which are used as inputs and not for evaluation.
This distinction is important to inform the reconstruction algorithm of the differences between events which are not encoded in the hit-based features $x$.

In the second step, the reconstruction is a second function $g:\theta, F_\text{sim} \mapsto E_\text{reco}$, that produces the reconstructed targets $E_\text{reco}$ containing all the high-level variables of interest based on the outputs of the simulation.
This is usually the end of the reconstruction chain, but in the present work we are interested in comparing the performance of different detector designs.
For this purpose, we compare $E_\text{true}$ and $E_\text{reco}$ using a Loss function, yielding a single scalar loss term $\mathcal{L}$ that describes the goodness of each detector design.
In essence, we can combine the effects of simulation and reconstruction on $\theta$ into a single function $V: \theta \mapsto \mathcal{L}$ that yields a unique Loss for each set of parameters $\theta$, excluding stochastic effects.
The function $V$ and by extension the Loss $\mathcal{L}$ include stochastic noise added by the simulation and potentially also by the reconstruction itself, {\it e.g.} if it is computed using a Deep Neural Network (DNN).
The stated goal of this work is then to efficiently explore the performance-space of $V: \theta \mapsto \mathcal{L}$ and find
\begin{equation}
    \theta_\text{opt} = \underset{\theta}{\text{argmin}} \left( \mathcal{L}, V \right).
    \label{eq:optimization_problem}
\end{equation}
By definition, $\theta_\text{opt}$ is the set of (optimal) parameters that minimizes $\mathcal{L}$.
This minimum can only be found using conventional gradient descent methods~\cite{ruder_overview_2017} as long as the stochastic noise of $\mathcal{L}$ is not too large.
In practice however, this noise tends to dominate the local Loss landscape, and methods such as the one described next are required to obtain meaningful gradients.

\section {Digital twin approach and optimization} \label{sec:digital_twin_approach}

In the previous Section, we introduced an example of a digital twin, namely the Geant4 simulation software that reproduces the behavior of a real-world detector.
This Section proposes a second digital twin to emulate the behavior of the combined simulation and reconstruction system.
This second digital twin is motivated by the fact that the loss $\mathcal{L}$ is both highly noisy and computationally expensive.
Indeed, even though a sequential pipeline that computes $V(\theta)=\mathcal{L}$ is sufficient for the task of optimization, it would require a full simulation and reconstruction from scratch for each gradient gradient step and be highly inefficient.
We demonstrate a significant improvement in computational efficiency by learning the expected detector performance with a Deep Neural Network, which addresses both challenges: first, once learned, its evaluation is inexpensive, and second, it partially smooths out irregularities in the gradient. 

\begin{figure}[htb]
    \centering
    \includegraphics[width=0.55\linewidth]{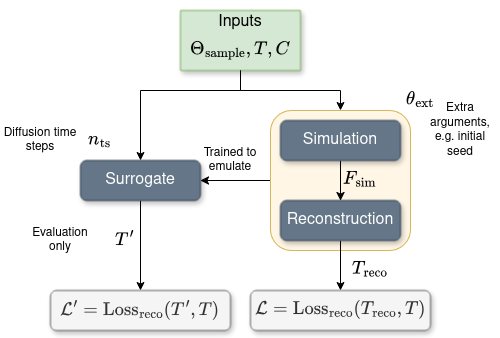}
    \caption{Illustration of the surrogate model as a digital twin of the combined reconstruction and simulation system. By avoiding the internal low-level information $F_\text{sim}$, the surrogate is able to efficiently produce similar outputs $T' \approx T_\text{reco}$.}
    \label{fig:digital_twin}
\end{figure}

This approach starts by sampling a subset $\Theta_\text{sample} \subseteq \Theta = \{ \theta_0, ..., \theta_N \}$ from a confined region within the larger parameter space.
We sample with a multi-variate normal distribution $\theta_\text{sample} = \mathcal{N}(\theta, K_\theta)$ for a total of $N$ simulations, where $K_\theta$ is the sampling covariance matrix.
A generative DNN, named the surrogate model in the following, learns the function $S$ that locally approximates the effects of $V(\theta)$.
In this way, the performance of a new parameter $\theta_{\text{new}} \notin \Theta_\text{sample}$ is approximated by evaluating $S(\theta_\text{new}) = \mathcal{L'_\text{new}}$, provided that $\theta_\text{new}$ is also within the training region $\Theta_\text{sample}$.
Fig. \ref{fig:digital_twin} shows this equivalence between the simulation-reconstruction system and the surrogate model.

The minimization problem stated in Eq.~\eqref{eq:optimization_problem} is performed locally by minimizing $\mathcal{L}'$ instead, which only requires the evaluation of $S(\theta)$. 
This approach enables the simulation to run in parallel, and the reconstruction algorithm to train on several detector designs at the same time, greatly reducing computation time.
To ensure that the learned approximation of $\mathcal{L}' \approx \mathcal{L}$ holds, it is necessary to validate the predictions of the surrogate model, as inaccuracies would lead to a misrepresentation of the physical detector.

\subsection{Surrogate model}
\label{sec:surrogate}
The surrogate model is a DNN tasked with reproducing the performance $\mathcal{L}'$ for any $\theta_\text{new}$ within the sampled region.
There are two main ways for the surrogate to learn $\mathcal{L}'$, either directly using $\mathcal{L}_\text{reco}$ or indirectly by learning $T_\text{reco}$ and evaluating $\mathcal{L}' = \text{Loss}_\text{reco}(T', T)$.
For this application, we train a conditional De-noising Diffusion Probabilistic model (DDPM)~\cite{ho_denoising_2020} on the targets $T_\text{reco}$.
Diffusion models are able to generate new data by training on progressively noisier samples and reversing the noise during evaluation.
Advantages of diffusion models include stable training, high fidelity samples and potentially easier transfer learning due to their probabilistic nature.

For our purposes, the internal training model is a simple feed-forward network composed of four linear layers, with the architecture detailed in Table~\ref{tab:surrogate_architecture}.
The training is done on the dataset $\{\theta, C, T_\text{reco}\}$ where the goal of the model is to predict the added noise.
We train the model using the Adam optimizer~\cite{kingma_adam_2017} and a Mean Squared Error (MSE) loss between the \textit{predicted} and the \textit{true} added noise.
The Adam optimizer is momentum-based and enhances the transfer learning between iterations. 
During evaluation, the surrogate predicts new targets $T' = S\left( \theta', C \right)$ used to approximate the reconstruction Loss $\mathcal{L'} = \text{Loss}_\text{reco}(T', T)$.
\begin{table}[ht]
    \centering
    \begin{tabular}{c c c c }
    \toprule
    \textbf{Layer} & \textbf{Input Nodes} & \textbf{Output Nodes} & \textbf{Activation Function} \\
    \midrule
    1              & $\theta + C + T_\text{reco}+ 1$& 200& ELU \\
    2              & 200& 100                  & ELU \\
    3              & 100                  & $T'$& Linear\\
    \bottomrule
\end{tabular}
\caption{Architecture of the DNN used by the surrogate model. The input nodes of the first layer are the detector parameters $\theta$, the context information $C$, the target quantities $T$ and the internal time step variable $n_\text{ts}$, which accounts for the $+1$ term.}
\label{tab:surrogate_architecture}
\end{table}

\subsection{Optimizer}

The optimizer object is a wrapper around the Adam optimizer that adjusts the parameters $\theta$ in order to minimize
\begin{equation}
    \mathcal{L}_\text{opt} = \mathcal{L}' + \mathcal{L}_\text{penalties} + \mathcal{L}_\text{boundaries},
\end{equation}
where $\mathcal{L}'$ is the Loss predicted by the surrogate, $\mathcal{L}_\text{penalties}$ is a user-defined additional term that takes into account effects such as total cost, maximum size or other physical constraints, and $\mathcal{L}_\text{boundaries}$ is a penalty term that ensures that the parameters remain within the evaluation space of the surrogate.
This boundary loss is defined as
\begin{equation}
    \mathcal{L}_\text{boundaries} = \text{mean} 
        \left(
              \frac{1}{2} 
              \left[ \text{Relu}
                \left(
                    \theta - \frac{\theta_\text{max}}{1.1}
                \right)
            \right]^2
            + \frac{1}{2} \left[ \text{Relu} \left(
                    \frac{\theta_\text{min}}{1.1} - \theta
            \right)
            \right] ^2
        \right).
\end{equation}
The scaling $1/1.1$ ensures that the penalty loss activates slightly before the actual limit is reached, allowing the optimizer to adjust to the penalty sooner.  
After each parameter update, we check that the parameters are still within the well-defined region of the surrogate evaluation space spanned by $K_\theta$ with the binary decision
\begin{equation}
    (\theta_{n+1} - \theta_n) K^{-1}_\theta (\theta_{n+1} - \theta_n) < b,
\end{equation}
where $\theta_{n+1}$ are the new parameters after the parameter update, $\theta_n$ are the central parameters of the trust region of the surrogate and the factor $b\leq 1$ is the scaling of the valid region for the surrogate.
We choose the conservative value $b=0.8$ to ensure that the optimized parameters $\theta_{n+1}$ remain well within the valid region.
As soon as this condition does not hold anymore, an early stopping activates, and the surrogate model is retrained on the new region.
This stopping defines the end of a single iteration of the simulation-reconstruction-surrogate-optimizer (SRSO) pipeline.
In order to encourage exploration in the next iteration, we adjust the sampling covariance $K_\theta$ given by
\begin{equation}
        K_\theta = \text{diag}(\sigma_0, ..., \sigma_m)^2 + (s - 1) \frac{(\theta_{n+1} - \theta_n) \otimes (\theta_{n+1} - \theta_n)}{|\theta_{n+1} - \theta_n|^2},
        \label{eq:covariance_matrix_adjustment}
\end{equation}
where $(\theta_{n+1} - \theta_{n})$ is the distance between the current and previous parameters and $s = 0.8 \cdot \text{max}(1, 4 |\theta_{n+1} - \theta_n|)$ is an appropriate scaling factor if each $\sigma_m \approx \mathcal{O}(1)$. 
Concretely, after each iteration where the optimizer reaches the boundaries, the sampling region expands in the direction of change.
The resulting modified parameters $\theta_\text{n+1}$ become the new central values for the next iteration.

\section {Workflow of the AIDO package} \label{sec:workflow}

\begin{figure}[htb]
    \centering
    \includegraphics[width=0.55\linewidth]{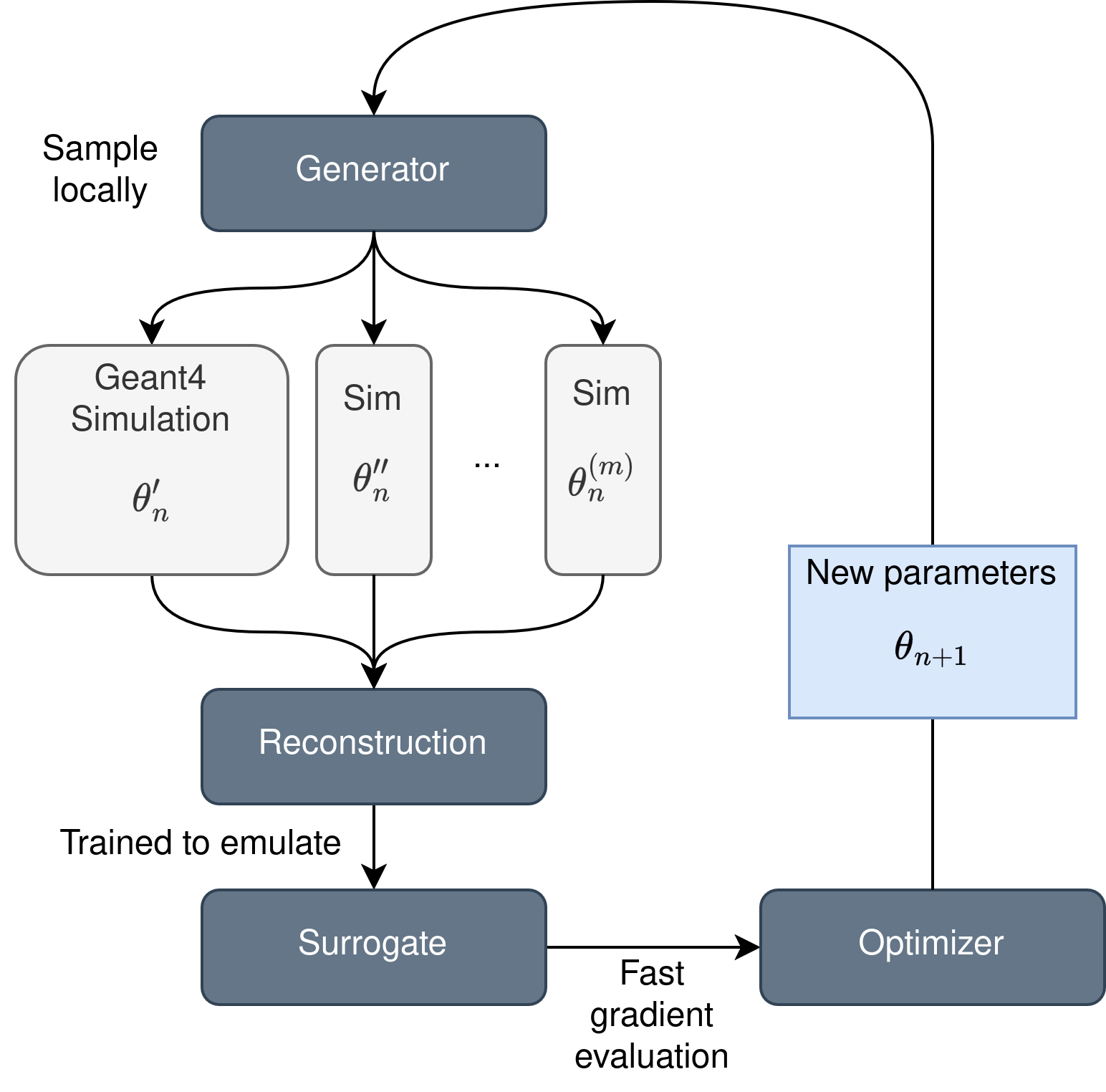}
    \caption{Workflow of a single SRSO iteration using the AIDO package. The top-level generator samples a new set of detector parameters which are simulated in parallel with Geant4. The outputs of the simulations are used to produce a single reconstruction $\mathcal{L}$ loss. Based on these, the surrogate and optimizer system predicts a new set of parameters for the next iteration.}
    \label{fig:aido_workflow}
\end{figure}

The AIDO software is a dedicated python package that performs the optimization of arbitrary detector hyperparameters as described in the previous Section.
The package is composed of two layers, first is a scheduler that coordinates the execution of the individual sub-components using b2luigi~\cite{michael_eliachevitch_belle2b2luigi_2025}, second is the implementation of the surrogate and optimizer models using pytorch~\cite{paszke_pytorch_2019}.
The AIDO scheduler subdivides the SRSO chain into independent Tasks that can be run on HPC using job schedulers such as HTCondor~\cite{htcondor_team_htcondor_2024}, as shown in Fig.~\ref{fig:aido_workflow}.

\subsection{Generation} \label{sec:generation}
We start with an initial set of parameters $\theta_\text{n=0}$, which can be chosen manually or randomly.
For a set of $N$ parallel simulations, we sample $N$ times from the multivariate normal distribution $\mathcal{N}\left( \theta, K_\theta \right)$, where the sampling covariance matrix $K_\theta$ encapsulates the correlations between each parameters.
This covariance is initialized as $K_\theta = \text{diag}\left( \sigma_0^2, ..., \sigma_m^2 \right)$, where $\sigma_m^2$ is the variance of each parameter.
The choice of the initial variance for each parameter is free, but is learned by the optimizer for all further iterations $n > 0$ according to Eq.~\eqref{eq:covariance_matrix_adjustment}. 
In addition, sampled parameters that lie outside the allowed boundaries are resampled until valid.
Each sampled $\theta^\text{sample}$ is then the configuration of a different detector, which are simulated in the next step.

% simulations (parallel, containerized)
\subsection{Simulation}
The simulations are executed in parallel, each with a unique set of detector parameters and a designated output file path.
In practice, the propagation of the parameters to Geant4 depends on the available interface.
By default, programs and end-users can communicate with a Geant4 executable by passing a macro file that lists a series of Geant4 commands.
The values of the detector hyperparameters can then be written with a single script to this file.
Some Geant4 programs are linked with a python interface (using C++ to python binding, such as pybind~\cite{wenzel_jakob_pybindpybind11_2024}).
This allows for a simple propagation of parameters to the individual simulation software.
Due to the external Geant4 libraries that are often required by individual Geant4 programs, the approach of containerization is highly recommended.
And while the output of Geant4 programs is in most cases a ROOT file~\cite{brun_root-projectroot_2019}, a specific file format is not required by the AIDO software.

% reconstruction (GPU, containerized)
\subsection{Reconstruction}
\label{sec:reconstruction}
Once all the simulations are run, the reconstruction computes the goodness of each design.
Since regular reconstruction algorithms (specially ML-based) are usually applied onto a single large dataset, the AIDO package provides a hook for merging and converting the output files of the simulations into a single dataset and writing it to file.
This step ensures that the reconstruction runs on a homogeneous dataset, with the correct format and within its own design conditions, providing maximal flexibility.
It is then the responsibility of the user to ensure that the reconstruction algorithm outputs a meaningful performance metric $\mathcal{L}$ for each design.

\section {Optimization of discrete parameters} \label{sec:discrete_parameters}

Categorical parameters are as such not suitable for optimization due to their inherent discontinuity.
To overcome this, instead of tuning a single discrete parameter, we assign a probability to each category, reflecting the model's confidence in that category.
Equivalently, we can sample from these probability distributions to produce a representative dataset.
The sampled categorical values are encoded as one-hot vectors in the training dataset of the surrogate.
A parameter $\phi$ composed of $i$ distinct categories $\phi = \{ \phi^0, ..., \phi^i \} $ is represented by the $i \times i$ matrix $M_\phi$, defined as
\begin{equation}
    M_\phi = \hat{\mathbbm{1}} = \begin{bmatrix}
        1       & 0         & \dots     & 0 \\
        0       & 1         & \dots     & 0 \\
        \vdots  & \vdots    & \ddots    & \vdots \\
        0       & 0         & \dots     & 1 \\
    \end{bmatrix},
\end{equation}
where the category $\phi^i$ is the $i$-th row vector.
The one-hot encoding scheme is paramount to ensure the proper communication between the surrogate and the optimizer.
While the surrogate is trained on a dataset composed of one-hot vectors, it is still able to interpret the query of the optimizer displayed as fractional values.
This is due to the inherent ability of DNNs to interpolate between values within their training space.
To facilitate training, instead of learning the probabilities directly, the optimizer adjusts the log-probabilities, or logits, $\{z_n\}$, which are defined as $P(\phi_n) = \text{softmax}(z_n)$.

\section{Application to a sampling calorimeter} \label{sec:application_to_sampling_calo}

As an illustration of possible applications for the AIDO framework, we showcase a sampling calorimeter composed of absorber and active scintillator layers.
We refer as absorber material to any passive component that is not part of the readout.
Scintillators on the other hand are materials that emit optical photons when they are traversed by high-energy particles, which allows for an electronic readout.
The layers themselves are shaped as rectangular cuboids with a side-length of \SI{50}{\cm} and are stacked longitudinally along the beam axis, alternating between absorber and scintillator.
We set the thickness along the beam axis and the material of the six layers as the optimizable parameters.
We bound the thickness of all layers to be strictly positive in order to avoid unphysical layouts.
The material is a binary choice between either lead or iron for each one of the absorber layers and between lead tungsten \ch{PbWO4} and polystyrene \ch{(C8H8)_n} for each scintillator layer.

The task of the optimizer is then to improve the energy resolution by adjusting the thickness of each layer and by choosing the suitable materials.
From a physical standpoint, a good calorimeter should maximize the amount of energy that get deposited by the particles as they travel through the material.
This can be done by increasing the volume of active material while balancing the volume of absorber, as passive material helps in slowing down highly energetic particles, making them more prone for interactions with the active material.
Based on the much shorter radiation length of \ch{PbWO4} (\SI{1.27}{\cm}) compared to polystyrene (\SI{41.31}{\cm}), we expect the optimizer to prefer the use of the former over the latter for the scintillator layers, with these values found in \cite{Workman:2022ynf}.
The choice of absorber is not as decisive, as the radiation lengths of \ch{Fe} (\SI{1.76}{\cm}) and \ch{Pb} (\SI{0.56}{\cm}) are similar.
In the absence of any other constraints, the optimal calorimeter for these conditions would be a single block of \ch{PbWO_4}.
To address practical considerations, we introduce cost and size constraints next.

\subsection{Additional constraints}

\begin{figure}[htb]
    \centering
    \includegraphics[width=0.65\linewidth]{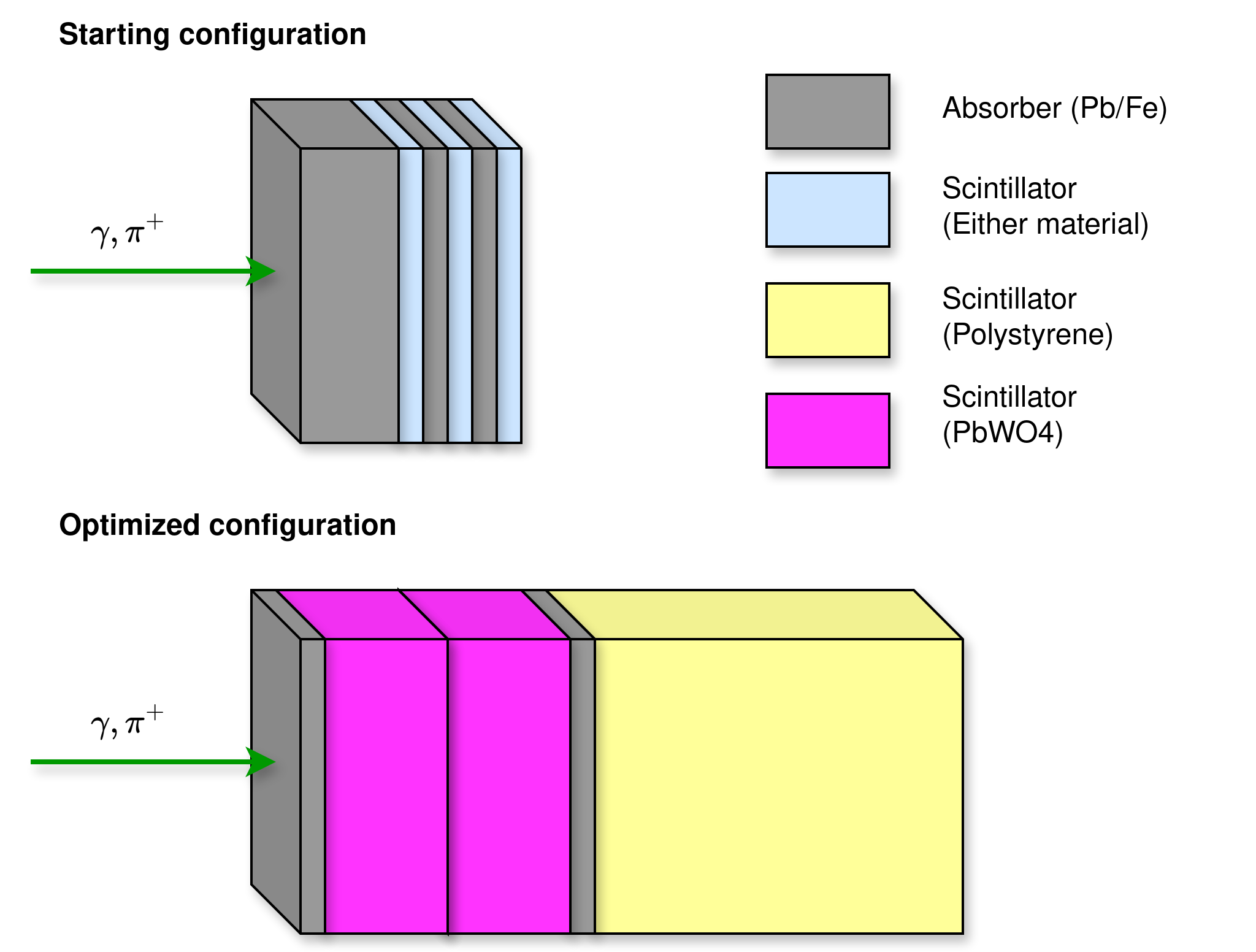}
    \caption{Detector configuration before optimization (upper) and after (lower). The incoming particles impinge perpendicularly on the detector. In the starting configuration, most particles are stopped by the absorber before they can be recorded by the scintillators. In the optimized configuration, photons are recorded by the large \ch{PbWO_4} block, while the longer pion showers are contained by the polystyrene section.}
    \label{fig:detector}
\end{figure}

In order to recreate a realistic engineering problem, we constrain the total length of the detector to $\text{d}_\text{max} = 200\,\si{\centi\m}$. This is enforced with a penalty term of the form
\begin{equation}
    \mathcal{L}_\text{length} = 10  \left( \text{Relu}( d - d_\text{max})\right)^2
\end{equation}
We set the maximal total cost for the whole detector at $\text{cost}_\text{max} = 200\text{k}\,\euro$, which is calculated based on the weighted cost for each layer
\begin{equation}
    \text{cost} = \sum_{l=0}^6 d_l c_l P(\phi_l)
\end{equation}
where $c_l$ is a vector of the same length as $\phi_l$ describing the cost for each category.
For the three absorber layers, the cost is $25\si{\europercm}$ for lead and $\SI{4.16}{\europercm}$ for iron, while for the scintillator layers the cost is $2.5\text{k}{\europercm}$ for \ch{PbWO_4} and $0.01{\europercm}$ for polystyrene.
We chose these values based approximately on the market price per volume, focusing on a large discrepancy between \ch{PbWO_4} and polystyrene.
In this way, we ask the optimizer to balance the performance gained by choosing the better material against the increase in total cost.
The penalty scaling for exceeding the allowed cost is
\begin{equation}
    \mathcal{L}_\text{cost} = \text{Relu}\left( \dfrac{\text{cost}}{\text{cost}_\text{max}}-1\right)^2.
\end{equation}
Finally, Fig.~\ref{fig:detector} shows the starting and the optimized configuration of the detector with the above cost and size constraints.

\subsection{Simulation and reconstruction}

For each iteration, we simulate $20$ distinct detector configurations, as described in Sec.~\ref{sec:generation}.
The simulations are run in parallel automatically by the AIDO scheduler, and each simulation runs in single-threaded mode with $400$ events.
In each event, we shoot a single initial particle, with a random energy uniformly sampled between $1$ and $20$ \GeV; in half of the events the particle is a photon, in the other half it is a positively charged pion. 
The goal of this example is to optimize the twelve detector parameters (layer thickness and material composition) such that we minimize the energy resolution of the total calorimeter.
For this purpose, we record the idealized total deposited energy in each of the three scintillator layers, without any readout effects.

\begin{table}[ht]
\centering
\begin{tabular}{c c c c }
\toprule
\textbf{Layer} & \textbf{Input Nodes} & \textbf{Output Nodes} & \textbf{Activation Function} \\
\midrule
1              & $\theta$& 100                  & ELU \\
2              & 100                  & 100                  & ELU \\
3              & 100                  & $F_{\text{sim}}$      & Relu \\
\bottomrule
\end{tabular}
\caption{Reconstruction model: pre-processing block architecture}
\label{tab:reco_pre_processing}
\end{table}

\begin{table}[ht]
\centering
\begin{tabular}{c c c c}
\toprule
\textbf{Layer} & \textbf{Input Nodes}             & \textbf{Output Nodes} & \textbf{Activation Function} \\
\midrule
1              & $\theta + F_{\text{sim}}$& 100                  & ELU \\
2              & 100                              & 100                  & ELU \\
3              & 100                              & 100                  & Relu \\
4              & 100                              & $T = 1$              & Linear \\
\bottomrule
\end{tabular}
\caption{Reconstruction model: main block architecture. We have one output node ($T=E_\text{rec}$), the reconstructed energy per event.}
\label{tab:reco_main_block}
\end{table}

The reconstruction algorithm is a feed-forward neural network implemented in pytorch, composed of two distinct blocks: a pre-processing block and a main block, summarized in Table~\ref{tab:reco_pre_processing} and Table~\ref{tab:reco_main_block} respectively.
The pre-processing block consists of three fully connected layers, designed to process the initial input parameters ($\#\theta=12$).
The main block processes is composed of four fully connected layers with a single output node being the reconstructed energy.
The learning is performed using the MSE loss between the reconstructed energy  $E_\text{rec}$ and the true initial energy  $E_\text{true}$ of the particle in each event,
\begin{equation}
        \mathcal{L} = \text{mean} \left( \dfrac{ \left( E_\text{rec} - E_\text{true} \right)^2}{E_\text{true} + 1} \right),
        \label{eq:reco_loss}
\end{equation}
where $E_\text{rec}$ is the predicted energy by the reconstruction model and $E_\text{true}$ the true MC initial energy.
We chose a scaling of $1/(E_\text{true}+1)$ to regulate the importance of events with higher initial energy.
In Sec.~\ref{sec:results}, we will discuss the impact of this Loss on the final detector configuration.
During the forward pass, the output of the pre-processing block is multiplied element-wise with the input vector $F_{\text{sim}}$.
The result is concatenated with the detector parameters $ \theta $ and passed into the main block.
For the surrogate and optimizer, we use the configuration detailed in Sec.~\ref{sec:digital_twin_approach}, with a learning rate of $\text{lr}_S = 0.001$ for the surrogate and $\text{lr}_O=0.01$ for the optimizer\footnote{All computations were performed on a system equipped with two Intel Xeon E5-2630 v4 processors and one NVIDIA Titan X GPU.}.

\subsection{Training validation}

To evaluate the performance of the reconstruction and surrogate models on unseen data, we generate a validation dataset under the same conditions as the training dataset, see Sec.~\ref{sec:generation}.
Before describing the validation results, it is important to emphasize that the goal of AIDO is to provide an algorithm capable of optimizing the detector design.
The use of a high-performing reconstruction algorithm is not the primary objective, as long as it performs adequately and consistently for all detector configurations.
Building on this premise, the validation of the reconstruction model is presented in addition to the surrogate model, which serves as a key step of the AIDO framework.

\begin{figure}[htb]
    \begin{subfigure}{0.49\textwidth}
        \centering
        \includegraphics[width=\linewidth]{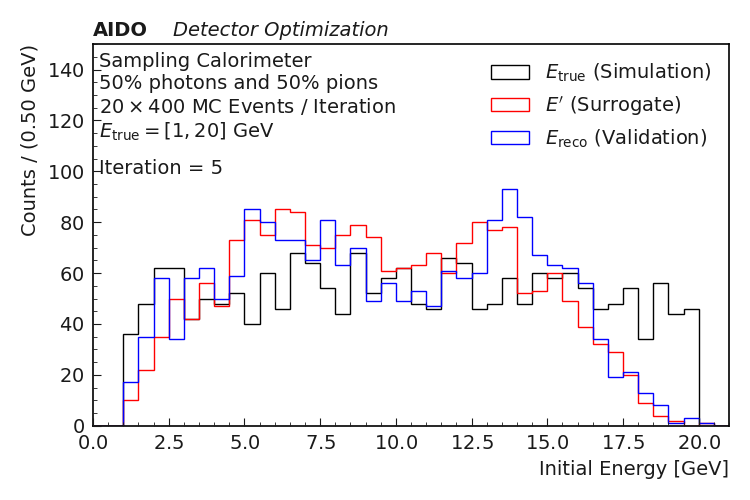}
        \caption{Validation before optimization.}
    \end{subfigure}
        \begin{subfigure}{0.49\textwidth}
        \centering
        \includegraphics[width=\linewidth]{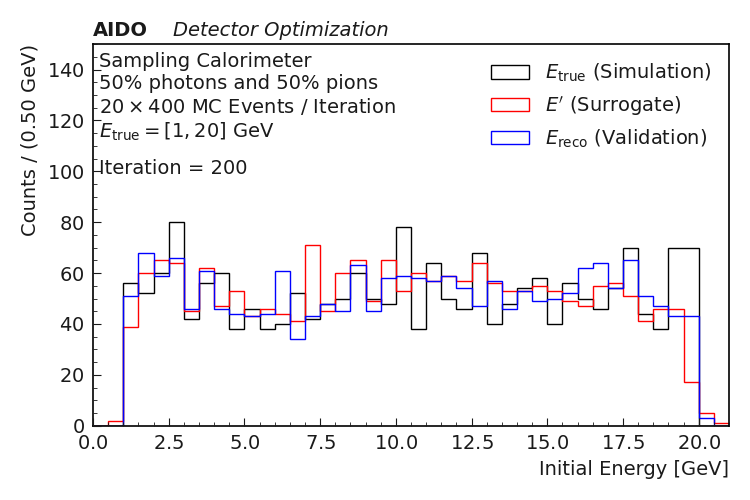}
        \caption{Validation after optimization.}
    \end{subfigure}
    \caption{Distribution of the initial energy of the primary particle as predicted by the reconstruction model ($E_\text{reco}$) and surrogate model ($E'$), along with the true simulated energy ($E_\text{true}$). The training was performed on $20 \times 400$ Monte Carlo events and the validation on $5 \times 400$ MC events. In both cases, the surrogate successfully emulates the distribution of samples produced by the reconstruction model.}
    \label{fig:validation}
\end{figure}

Figure~\ref{fig:validation} compares the distribution of the simulated energy with the energy predicted by the reconstruction and the surrogate models.
The reconstruction model reproduces the overall distribution of the targets and the surrogate model generates a similarly distributed sample over the range of targets.
As already pointed out in Ref.~\cite{Shirobokov:2020tjt}, the surrogate only needs to provide a reasonable gradient estimate and does not need to model specific details in most cases. 

\subsection{Results} 
\label{sec:results}

\begin{figure}[htbp]
    \centering
    \includegraphics[width=0.7\textwidth]{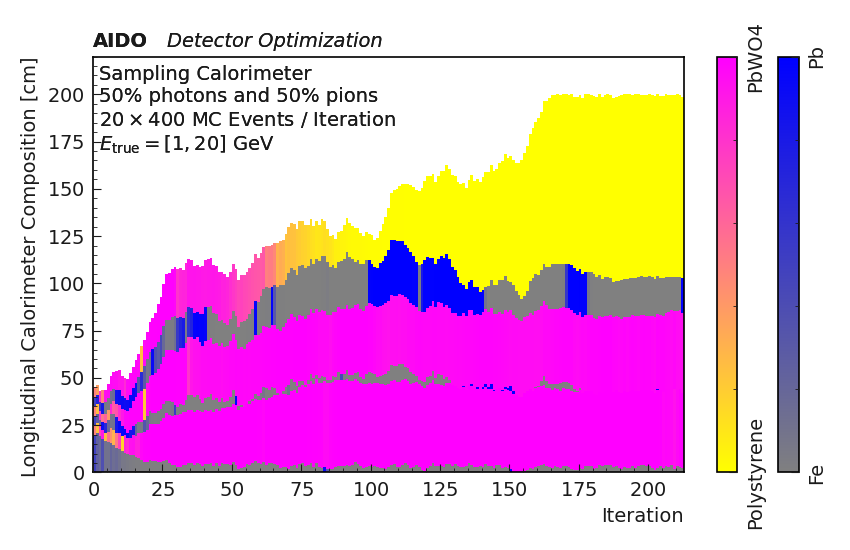}
    \caption{Sliced view of the longitudinal detector composition, with the initial setup at $\text{iteration}=0$, and the optimized setup at $\text{iteration}=220$. Each slice on the $x$-axis represents the current learned detector configuration (with the side at $y=0$ facing in the direction of the incoming particle beam, as shown in Fig.~\ref{fig:detector}). During training, the large block of the first absorber is gradually thinned while the scintillator layers grow in size. % In addition, the material choice for the scintillators settles on \ch{PbWO_4}.
    }
    \label{fig:calo_side_view}
\end{figure}

\begin{figure}[htbp]
    \centering
    \includegraphics[width=0.65\textwidth]{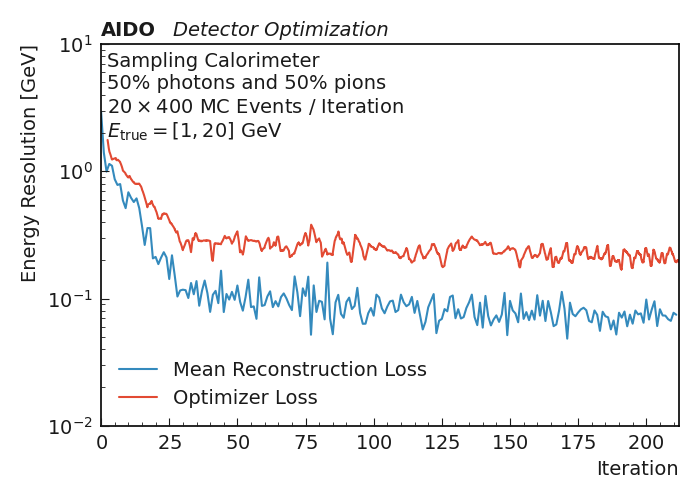}
    \caption{Evolution of the mean reconstruction loss $\mathcal{L}$ for the nominal design during training, as given by Eq.~\eqref{eq:reco_loss}, and the synthetic reconstruction loss $\mathcal{L}'$ used as the optimizer loss.}
    \label{fig:energy_resolution_evolution}
\end{figure}

The evolution of the detector composition is shown in Fig.~\ref{fig:calo_side_view} with the corresponding optimizer loss displayed in Fig.~\ref{fig:energy_resolution_evolution}.
From a pure machine learning standpoint, the demonstration is successful: the target metric decreases in a reliable way, confirming the capability of the model to solve the optimization task.
However, we can gain further insights into the decisions taken by the model with some task-specific knowledge.

\begin{figure}[htb]
    \begin{subfigure}{0.49\textwidth}
        \centering
        \includegraphics[width=\linewidth]{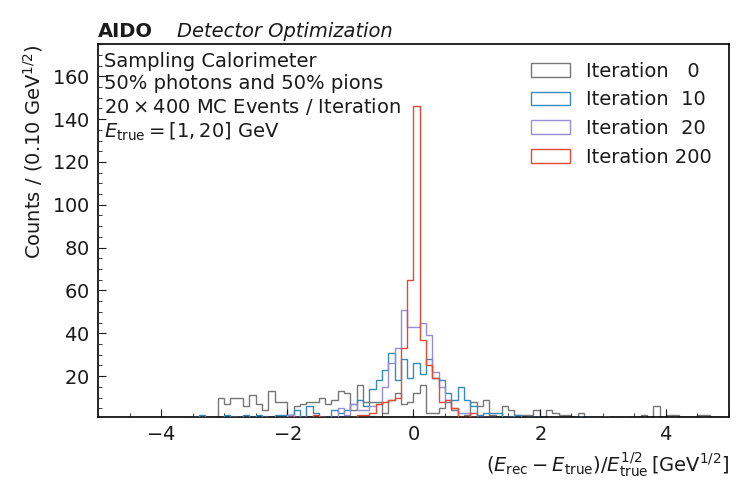}
        \caption{Energy resolution scaled with $1/\sqrt{E_\text{true}}$.}
    \end{subfigure}
    \begin{subfigure}{0.49\textwidth}
        \centering
        \includegraphics[width=\linewidth]{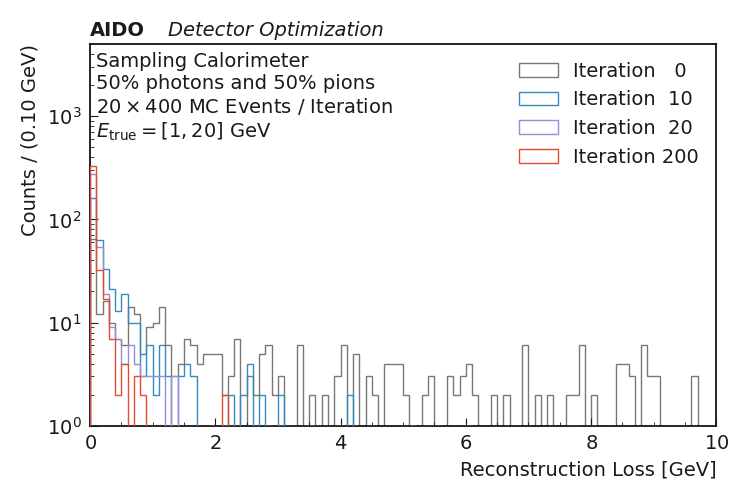}
        \caption{Reconstruction Loss.}
    \end{subfigure}
    \caption{Scaled energy resolution (a) and reconstruction loss (b) of the nominal training dataset ($400$ MC events) for selected iterations. The narrowing of both distributions indicate the gradual improvement achieved by the network.}
    \label{fig:energy_resolution_all}
\end{figure}

The initial setup has a large absorber block at the front which traps most of the particles before they can reach the first scintillator layer. 
Only a few particles are recorded by the subsequent scintillator, so that only a small fraction of the training dataset carries meaningful information.
In turn, the reconstruction performance is poor, the energy resolution is large and therefore the optimizer loss is high.

During training, the undesirable absorber layer is gradually removed, which has a net positive impact on the target energy resolution, as shown in Fig.~\ref{fig:energy_resolution_all}.
% It remains as a thin layer due to the $1/(E_\text{true})$ factor in the reconstruction loss, which encourages the optimizer to absorb very low-energy events.
At the same time, the scintillator layers increase in length, and since the energy resolution of a calorimeter scales with $1/\sqrt{E}$~\cite{Workman:2022ynf}, a larger active volume leads directly to more deposited energy and better energy resolution.
As expected, the optimizer chooses \ch{PbWO_4} over polystyrene for the first layer.
This makes sense for measuring the energy of photons, as electromagnetic showers benefit from a short but dense active material.

\begin{figure}
    \centering
    \includegraphics[width=0.65\linewidth]{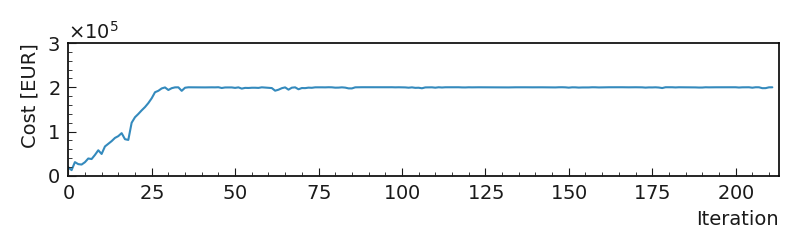}
    \caption{Evolution of the cost constraints during training. The initial growth is mainly due to the increase in thickness of the scintillator layers made out of the more expensive \ch{PbWO_4}.}
    \label{fig:constraints}
\end{figure}

After this electromagnetic section, the calorimeter displays a long heterogeneous part. Given the radiation length of \ch{PbWO_4}, electromagnetic showers are not expected to have a longitudinal profile that significantly extends beyond $25$ \si{\centi\m} of material. However, also for early showering charged pions, \ch{PbWO_4} shows excellent properties. Therefore, as can be seen in Fig.~\ref{fig:constraints}, the available budget is fully exploited, with a focus on the earlier layers, with very little or no absorber material.

\begin{figure}[p]
    \centering
    \begin{minipage}{0.4\textwidth}
        \includegraphics[width=\textwidth]{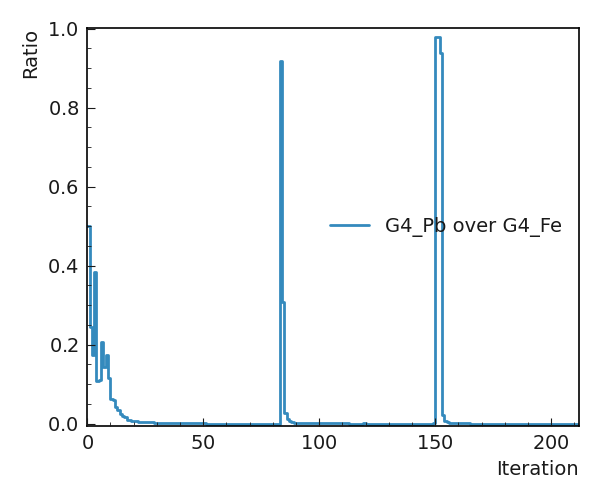}
        \caption*{(a) Absorber layer 0 \\(Pb or Fe)}
    \end{minipage}
    \begin{minipage}{0.4\textwidth}
        \centering
        \includegraphics[width=\textwidth]{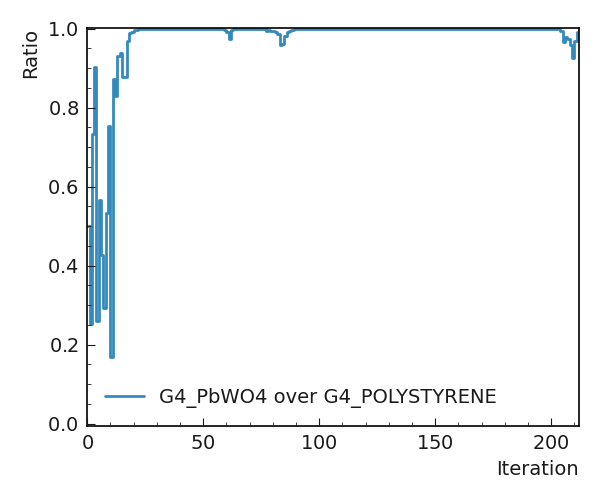}
        \caption*{(b) Scintillator layer 0 \\(PbWO4 or Polystyrene)}
    \end{minipage}
    \vspace{0.5cm}

    % Row 2
    \begin{minipage}{0.4\textwidth}
        \centering
        \includegraphics[width=\textwidth]{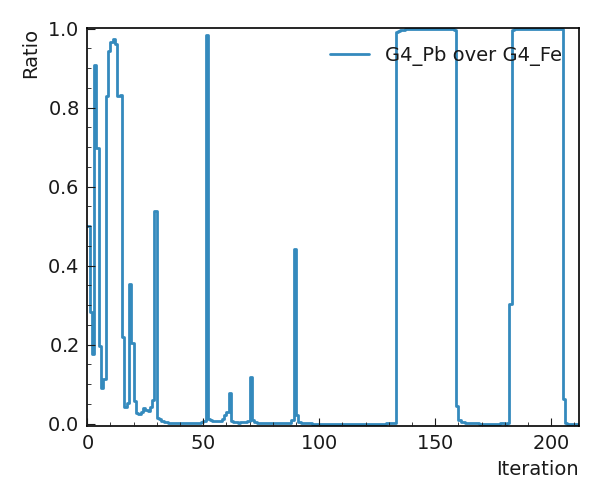}
        \caption*{(c) Absorber layer 1 \\(Pb or Fe)}
    \end{minipage}
    \begin{minipage}{0.4\textwidth}
        \centering
        \includegraphics[width=\textwidth]{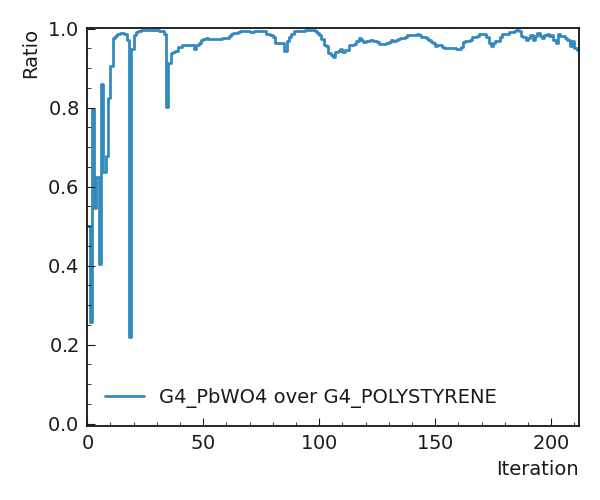}
        \caption*{(d) Scintillator layer 1 \\(PbWO4 or Polystyrene)}
    \end{minipage}
    \vspace{0.5cm}

    % Row 3
    \begin{minipage}{0.4\textwidth}
        \centering
        \includegraphics[width=\textwidth]{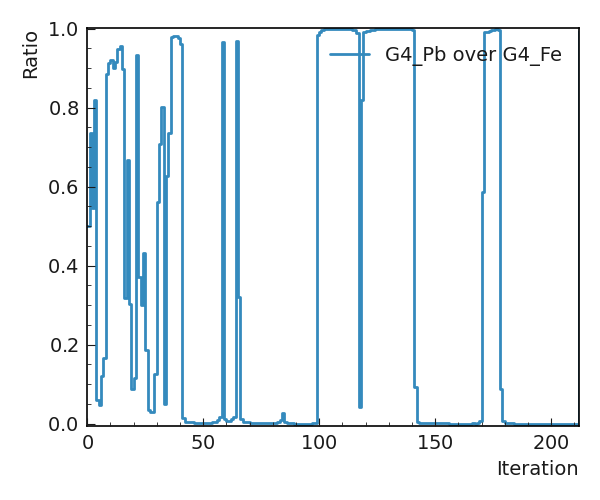}
        \caption*{(e) Absorber layer 2 \\(Pb or Fe)}
    \end{minipage}
    \begin{minipage}{0.4\textwidth}
        \centering
        \includegraphics[width=\textwidth]{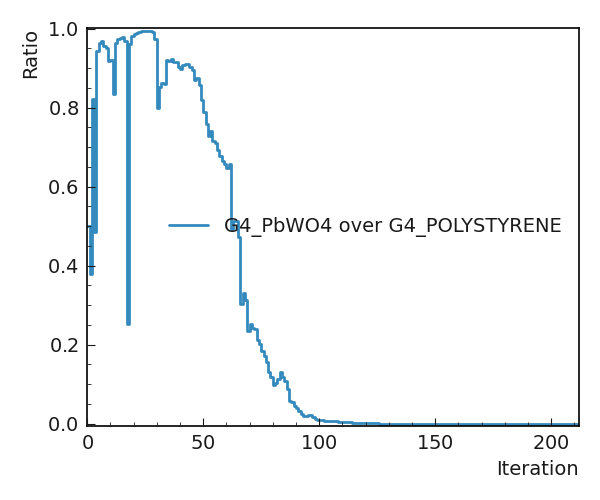}
        \caption*{(f) Scintillator layer 2 \\(PbWO4 or Polystyrene)}
    \end{minipage}
    \vspace{0.5cm}

    \caption{Evolution of the confidence of the optimizer in each material choice. The figures show the fraction of material proposed by the optimizer, with high fractions being considered more desirable by the optimizer. Crucially, the optimizer correctly chooses \ch{PbWO_4} as the material for the first and second scintillator layers (b) and (d). The third layer (f) also start with \ch{PbWO_4} but revert to polystyrene once the cost constraints apply, in order to save cost.}
    \label{fig:probabilities_over_time}
\end{figure}

Once the current composition exhausts the total available budget, the last layer changes to a less expensive material with large radiation length and therefore requires an absorber to shower the particles consistently.
The large iron absorber and the polystyrene scintillator build the last layer, with the shift from \ch{PbWO_4} to polystyrene clearly shown in Fig.~\ref{fig:probabilities_over_time}.
This last section of the detector helps by measuring high-energy pions that shower in the iron block, leaving a meaningful signal in the polystyrene downstream.
It can be concluded that the method is able to reconcile the multi-target optimization, taking into account physics performance as well as cost and length constraints.
In particular, the system is capable of optimizing discrete material choices with vastly different cost. 

\section {Conclusion} \label{sec:conclusion}

In this work, we demonstrate a novel method for investigating the configuration phase space of modern detectors.
By using a diffusion-based surrogate model that locally interpolates the behavior of the detector, we are able to efficiently explore and optimize this configuration space.
The method also extends nicely to categorical parameters such as the material choice, by adjusting the log-probabilities associated with each category.
The AIDO python package allows for large scale deployment on HPC facilities of the optimization pipeline by atomizing it into separate luigi Tasks.
We confirm the capabilities of the surrogate approach by optimizing a layered calorimeter, which converges towards a familiar design with an electromagnetic section, a part for early showering pions, and a simple sampling calorimeter capturing the remaining hadrons.
In the future, our aim will be to investigate further the performance of this model for more complex tasks, including larger parameter spaces and other performance metrics, such as the accuracy of tracking modules.
In principle, this method and framework are not limited to calorimeters, such that other tasks including tracking can be considered in the future.

\dataavailability{The resources used for the analysis can be found on the following GitHub page: \url{https://github.com/KylianSchmidt/aido}.}

\funding{
The work by TD and FS was partially supported by the Wallenberg AI, Autonomous Systems and Software 
Program (WASP) funded by the Knut and Alice Wallenberg Foundation.
The work by MA and FS was partially supported by the Jubilee Fund at the
 Lule{\aa} a University of Technology.
The work by PV was supported by the ``Ramón y Cajal” program under Project No. RYC2021-033305-I funded by the MCIN MCIN/AEI/10.13039/501100011033 and by the European Union NextGenerationEU/PRTR.
JK is supported by the Alexander-von-Humboldt foundation.
}

\abbreviations{Abbreviations}{
The following abbreviations are used in this manuscript:\\
\noindent 
\begin{tabular}{@{}ll}
 MDPI & Multidisciplinary Digital Publishing Institute\\
 SRSO & Simulation-Reconstruction-Surrogate-Optimizer\\
 DNN & Deep Neural Network\\
 DDPM & De-noising Diffusion Probabilistic model\\
 HPC & High Performance Computing\\
 HEP & High Energy Physics\\
 MSE & Mean Square Error\\
 ELU & Exponential Linear Unit\\
\end{tabular}
}

%%%%%%%%%%%%%%%%%%%%%%%%%%%%%%%%%%%%%%%%%%
%% Optional
\appendixtitles{no} % Leave argument "no" if all appendix headings stay EMPTY (then no dot is printed after "Appendix A"). If the appendix sections contain a heading then change the argument to "yes".
\appendixstart
\appendix

%%%%%%%%%%%%%%%%%%%%%%%%%%%%%%%%%%%%%%%%%%
\begin{adjustwidth}{-\extralength}{0cm}
%\printendnotes[custom] % Un-comment to print a list of endnotes

\reftitle{References}

% Please provide either the correct journal abbreviation (e.g. according to the “List of Title Word Abbreviations” http://www.issn.org/services/online-services/access-to-the-ltwa/) or the full name of the journal.
% Citations and References in Supplementary files are permitted provided that they also appear in the reference list here. 

%=====================================
% References, variant A: external bibliography
%=====================================
\bibliography{main}

% If authors have biography, please use the format below
%\section*{Short Biography of Authors}
%\bio
%{\raisebox{-0.35cm}{\includegraphics[width=3.5cm,height=5.3cm,clip,keepaspectratio]{author1.pdf}}}
%{\textbf{Firstname Lastname} Biography of first author}
%
%\bio
%{\raisebox{-0.35cm}{\includegraphics[width=3.5cm,height=5.3cm,clip,keepaspectratio]{author2.jpg}}}
%{\textbf{Firstname Lastname} Biography of second author}

% For the MDPI journals use author-date citation, please follow the formatting guidelines on http://www.mdpi.com/authors/references
% To cite two works by the same author: \citeauthor{ref-journal-1a} (\citeyear{ref-journal-1a}, \citeyear{ref-journal-1b}). This produces: Whittaker (1967, 1975)
% To cite two works by the same author with specific pages: \citeauthor{ref-journal-3a} (\citeyear{ref-journal-3a}, p. 328; \citeyear{ref-journal-3b}, p.475). This produces: Wong (1999, p. 328; 2000, p. 475)

%%%%%%%%%%%%%%%%%%%%%%%%%%%%%%%%%%%%%%%%%%
%% for journal Sci
%\reviewreports{\\
%Reviewer 1 comments and authors’ response\\
%Reviewer 2 comments and authors’ response\\
%Reviewer 3 comments and authors’ response
%}
%%%%%%%%%%%%%%%%%%%%%%%%%%%%%%%%%%%%%%%%%%
\PublishersNote{}
\end{adjustwidth}
\end{document}